\documentclass[aps,prb,twocolumn,superscriptaddress,showpacs]{revtex4-1}
\usepackage[english]{babel}
\usepackage[T1]{fontenc}
\usepackage[utf8]{inputenc}
\usepackage{lmodern}
\usepackage{amsmath,amssymb,graphicx,xcolor,bm,soul,url,hyperref}
\hypersetup{colorlinks=true,citecolor={blue},linkcolor={blue},urlcolor={blue}}
\usepackage{natbib}

\begin{document}

\title{Spontaneous and Superfluid Chiral Edge States in Exciton-Polariton Condensates}

\author{H. Sigurdsson}
\email[correspondence address:~]{helg@hi.is}
\affiliation{Science Institute, University of Iceland, Dunhagi-3, IS-107 Reykjavik, Iceland}

\author{G. Li}
\affiliation{School of Physics and Astronomy, University of Southampton, Southampton SO17 1BJ, United Kingdom}

\author{T. C. H. Liew}
\affiliation{Division of Physics and Applied Physics, School of Physical and Mathematical Sciences, Nanyang Technological University, 21 Nanyang Link, Singapore 637371}

\begin{abstract}
We present a scheme of interaction-induced topological bandstructures based on the spin anisotropy of exciton-polaritons in semiconductor microcavities. We predict theoretically that this scheme allows the engineering of topological gaps, without requiring a magnetic field or strong spin-orbit interaction (transverse electric-transverse magnetic splitting). Under non-resonant pumping, we find that an initially topologically trivial system undergoes a topological transition upon the spontaneous breaking of phase symmetry associated with polariton condensation. Under resonant coherent pumping, we find that it is also possible to engineer a topological dispersion that is linear in wavevector -- a property associated with polariton superfluidity.
\end{abstract}

\date{\today}

\maketitle

The hybridization of light and matter in the form of exciton-polaritons in microcavities has led to a new kind of quantum fluid~\cite{Carusotto2013}, well-known for its ability to develop coherence spontaneously as a Bose-Einstein condensate~\cite{Kasprzak2006,Byrnes2014} and flow without friction as a superfluid~\cite{Amo2009,Sanvitto2010}, even at room temperature~\cite{Lerario2017}. Furthermore it has been shown that exciton-polaritons can be manipulated by highly tuneable optically-induced potentials~\cite{Amo2010,Wertz2012}. These allow introducing a spatial structure in an otherwise homogeneous system, which has given access to a variety of fundamental effects, including: the gating~\cite{Gao2012,Anton2013} and routing~\cite{Flayac2013,Marsault2015} of polariton flow; the trapping of polariton superfluids~\cite{Tosi2012,Cristofolini2013}; the formation of patterns~\cite{Manni2011}; the breaking of chiral symmetry~\cite{Dall2014}; and the exposure of exceptional points~\cite{Gao2015}.

Taking inspiration from the field of topological photonics~\cite{Lu2014}, recent theoretical works have considered engineering topological polariton bandstructures~\cite{Karzig2015,Bardyn2015,Nalitov2015,Yi2016}. These are characterized by the formation of chiral edge states, at the boundaries between areas with different topology, which exhibit uni-directional propagation and an absence of backscattering even in the presence of defects or disorder. Due to these properties, chiral edge states are highly relevant to the field of polaritonics, which seeks robust mechanisms of propagating fields between individual information processing elements to allow for cascadable systems~\cite{Ballarini2013,Sanvitto2016}. The presence of Kerr-type interactions between polaritons would also allow the development of a nonlinear topological photonics, where schemes for solitons forming in the chiral edge modes have also appeared in recent theoretical works~\cite{Leykam2016,Kartashov2016,Gulevich2017}.

The previous schemes of topological polaritons have been based on three ingredients. First, a periodic potential is required to introduce the bandstructure on which to impose non-trivial topology. This has implicated the need for hard engineering of the polariton potential, such as is achieved by etching micropillar arrays~\cite{Milicevic2015}. Second, so that edge states propagate in only one direction, time-reversal symmetry should be broken, which implies the application of a magnetic field. Third, a significant amount of spin-orbit coupling is needed, which implies strong transverse electric-transverse magnetic (TE-TM) splitting. While large TE-TM splitting (on the order of tenths of milli-electronvolts) was achieved in samples from decades ago~\cite{Panzarini1999}, microcavities have evolved over the years to reduce this splitting (the quality factor is optimum when the cavity mode frequency is at the center of the stop-band, where the TE-TM splitting is smallest). Even though the aforementioned three ingredients can be achieved in principle, we will show that actually none of them are essential for generating topological polaritons!

We will consider a planar microcavity (with no etching), corresponding to an initially topologically trivial system, and illumination by a spatially patterned optical field, strong enough to place polaritons in a nonlinear regime. We find that the spin-anisotropy of polariton-polariton interactions induces an effective spin-orbit coupling and breaking of time-reversal symmetry. Remarkably, since these effects depend on the polariton interaction energy (blueshift), we obtain topological gaps that exceed typical strengths of disorder. The effect is readily observable making use of polarization filtering to separate the injected condensate and the topological behaviour.

Aside the interest that topological photonics brings to polaritonics, there is also a question of whether or not exciton-polaritons can bring anything new to the field of topological photonics? Here we find that the specific features of polaritons play different roles depending on the excitation conditions. We consider both resonant and non-resonant pumping -- the two most commonly used schemes for exciting polaritons. Under non-resonant pumping, polaritons must achieve their coherence spontaneously. Consequently the topological behaviour that arises from an initially trivial system is also developed spontaneously, where a random choice of the system chirality determines in which way chiral edge states will propagate. In the case of resonant pumping, the combination of topological dispersion with an additional potential allows modification to a dispersion linear in the wavevector, which is generally considered a sufficient condition for polariton superfluidity~\cite{Carusotto2004,Shelykh2006,Carusotto2013}. As far as we know, these are unique features that do not appear in other topological photonic systems.

{\bf Theoretical Model.---} We begin by defining the wavefunctions of polaritons in the $x$ and $y$ linear in-plane polarizations as $\psi_x$ and $\psi_y$, respectively. Their evolution is determined by the polarization dependent driven-dissipative Gross-Pitaevskii equations~\cite{Shelykh2006}:
\begin{align}
i\hbar\frac{\partial\psi_{x,y}}{\partial t}&=\Big(E_{x,y}-\frac{\hbar^2}{2m}\nabla^2-\frac{i\Gamma_{x,y}}{2}+V(\mathbf{x})+iP_(\mathbf{x})\notag\\
&\hspace{10mm}+\left(U_0-i\Gamma_\mathrm{NL}\right)\left(|\psi_x|^2+|\psi_y|^2\right)\Big)\psi_{x,y}\notag\\
&\hspace{3mm}-U_1\left(|\psi_{x,y}|^2\psi_{x,y}+\psi_{y,x}^2\psi_{x,y}^*\right)+F_{x,y}(\mathbf{x})e^{-i\omega_pt}\label{eq:GP}
\end{align}
Here the equations are written in a general form, allowing for different energies (polarization splitting) of the $x$ and $y$ components, $E_x$ and $E_y$, respectively, and different lifetimes, $\Gamma_x$ and $\Gamma_y$, respectively. The effective polariton mass $m$ and potential $V(\mathbf{x})$ are polarization independent. We account for the potential term here for generality, being interested in both spatially homogeneous and etched microcavities. The nonlinear polariton-polariton interaction constants are related to those in the spinor basis~\cite{Shelykh2010} by $U_0=\alpha_1$ and $U_1=(\alpha_1-\alpha_2)/2$. It is well established that typically $\alpha_2$ is negative, while $\alpha_1$ is positive~\cite{Krizhanovskii2006,Leyder2007}. The nonlinear loss terms $\Gamma_\mathrm{NL}$ are most relevant when considering non-resonant pumping~\cite{Keeling2008}. Experiments on spin bifurcations have been fitted assuming that the nonlinear loss rate is polarization independent and that $\Gamma_x\neq\Gamma_y$~\cite{Ohadi2015,Ohadi2016}.

{\bf Excitation Schemes.---} In the following, we will consider different mechanisms of driving the system: 1) non-resonant excitation, for which $F_{x,y}=0$ and the potential term should be supplemented by a pump-induced shift~\cite{Wouters2007}, $V(\mathbf{x})=gP(\mathbf{x})$, where $g$ is a dimensionless constant; and 2) resonant $x$-linearly polarized excitation, for which $\Gamma_{NL}\approx0$, $P(\mathbf{x})=0$, $F_y=0$, and $\omega_p$ is the pump frequency. In either case we can expect the polariton condensate to be polarized in the $x$-direction. Under resonant pump this is obvious due to direct injection of $x$-polarized polaritons. Under non-resonant pumping, polariton condensates are also typically linearly polarized in microcavities where there is a polarization splitting~\cite{Kasprzak2007}. This is modelled by allowing $\Gamma_x\neq \Gamma_y$.

For simplicity, we now consider the case $U_0=U_1=\alpha$, which is consistent with experimental measurements in Refs.~\cite{Vladimirova2010,Takemura2014}. In this limit driving of the $x$-polarization does not excite significantly the $y$-polarization. Setting, $\psi_y=0$, decouples the equation for evolution of the $x$-polarized component:
\begin{align}
i\hbar\frac{\partial\psi_x}{\partial t}&=\Big(E_x-\frac{\hbar^2}{2m}\nabla^2-\frac{i\Gamma_x}{2}+V(\mathbf{x})+iP(\mathbf{x})\notag\\
&\hspace{15mm}-i\Gamma_\mathrm{NL}|\psi_x|^2\Big)\psi_x+F_x(\mathbf{x})e^{-i\omega_pt}\label{eq:GPx}
\end{align}
Note that the $U_0$ and $U_1$ dependent terms involving $|\psi_x|^2$ have been canceled.

We will take the driving fields to have a lattice type structure, where $F_{x,y}(\mathbf{x})$ and $P_{x,y}(\mathbf{x})$ are periodic in the $x$-direction with periodicity $a$. It is then appropriate to take $a$ as the natural unit of length and $\epsilon=\hbar^2/(2ma^2)$ as the natural unit of energy. With these choices, Eq.~\ref{eq:GPx} can be rescaled to take the form:
\begin{align}
i\frac{\partial\psi'_x}{\partial t'}&=\Big(E_x'-\hbar\omega_p'-\nabla'^2-\frac{i\Gamma_x'}{2}+V'(\mathbf{x})+iP'(\mathbf{x})\notag\\
&\hspace{15mm}-i|\psi_x'|^2\Big)\psi_x'+F_x'(\mathbf{x})\label{eq:GPxscaled}
\end{align}
where $\psi_x'=\sqrt{\Gamma_\mathrm{NL}/\epsilon}\psi_x\exp(-i\omega_pt)$, $t'=\epsilon t/\hbar$, $x'=x/a$, $y'=y/a$, $F_x'=F_x\sqrt{\Gamma_\mathrm{NL}}/\epsilon^{3/2}$, and all other primed parameters are scaled by $\epsilon$ (e.g., $E_{x,y}'=E_{x,y}/\epsilon$).

{\bf Dispersion.---} While the effective mass characterizes the dispersion of polaritons in the low-density regime, under the build-up of significant polariton populations the dispersion becomes renormalized~\cite{Carusotto2004,Utsunomiya2008,Kohnle2011,Pieczarka2015}. Let us focus our attention on the dispersion of $y$-polarized polaritons, which evolve under the scaled evolution equation:
\begin{align}
i\frac{\partial\psi'_y}{\partial t'}&=\Big(E_y'-\hbar\omega_p'-\nabla'^2-\frac{i\Gamma_y'}{2}+V'(\mathbf{x})+iP'(\mathbf{x})\notag\\
&\hspace{15mm}-i|\psi_x'|^2+\alpha'|\psi_x'|^2\Big)\psi_y'-\alpha'\psi_x^{\prime2}\psi_y^{\prime*}\label{eq:GPyscaled}
\end{align}
where $\alpha'|\psi_x'|^2=\alpha|\psi_x|^2/\epsilon$ and we continue working in the regime where the occupation of $\psi_y$ is small.

Due to the last term in Eq.~\ref{eq:GPyscaled} the dispersion of the $y$-polarized polaritons should be calculated by substituting a form $\psi_y'=A_{k_x}(\mathbf{x})e^{i\omega t}+B_{k_x}(\mathbf{x})e^{-i\omega^* t}$, where $A_{k_x}(\mathbf{x})$ and $B_{k_x}(\mathbf{x})$ are the wavevector dependent Bloch envelope functions accounting for the periodicity $a$. The positive and negative frequency components are coupled upon substitution into Eq.~\ref{eq:GPyscaled} and we obtain two coupled equations, which can be written in matrix form:
\begin{equation}
\left(\begin{array}{cc}E'&-\alpha'\psi_x^{\prime2}\\ \alpha'\psi_x^{\prime*2}&-E^{\prime*}\end{array}\right)\left(\begin{array}{c}A_{k_x}(\mathbf{x})\\B_{k_x}(\mathbf{x})\end{array}\right)=\hbar\omega'\left(\begin{array}{c}A_{k_x}(\mathbf{x})\\B_{k_x}(\mathbf{x})\end{array}\right)\label{eq:Bogoliubov}
\end{equation}
where:
\begin{align}
E'&=-\nabla'^2+E_y'-E_p'-\frac{i\Gamma_y'}{2}+(\alpha'-i\Gamma'_\mathrm{NL})|\psi_x'|^2\notag\\
&\hspace{15mm}+V'(\mathbf{x})+iP(\mathbf{x})'
\end{align}

The eigenvalues $\hbar\omega'$ of Eq.~\ref{eq:Bogoliubov} determine the energy spectrum. Having written the equations for generic driving of the system, either non-resonant or resonant, we will focus separately on these two cases in the following.

\begin{figure}[!t]
  \centering
	\includegraphics[width=0.9\columnwidth]{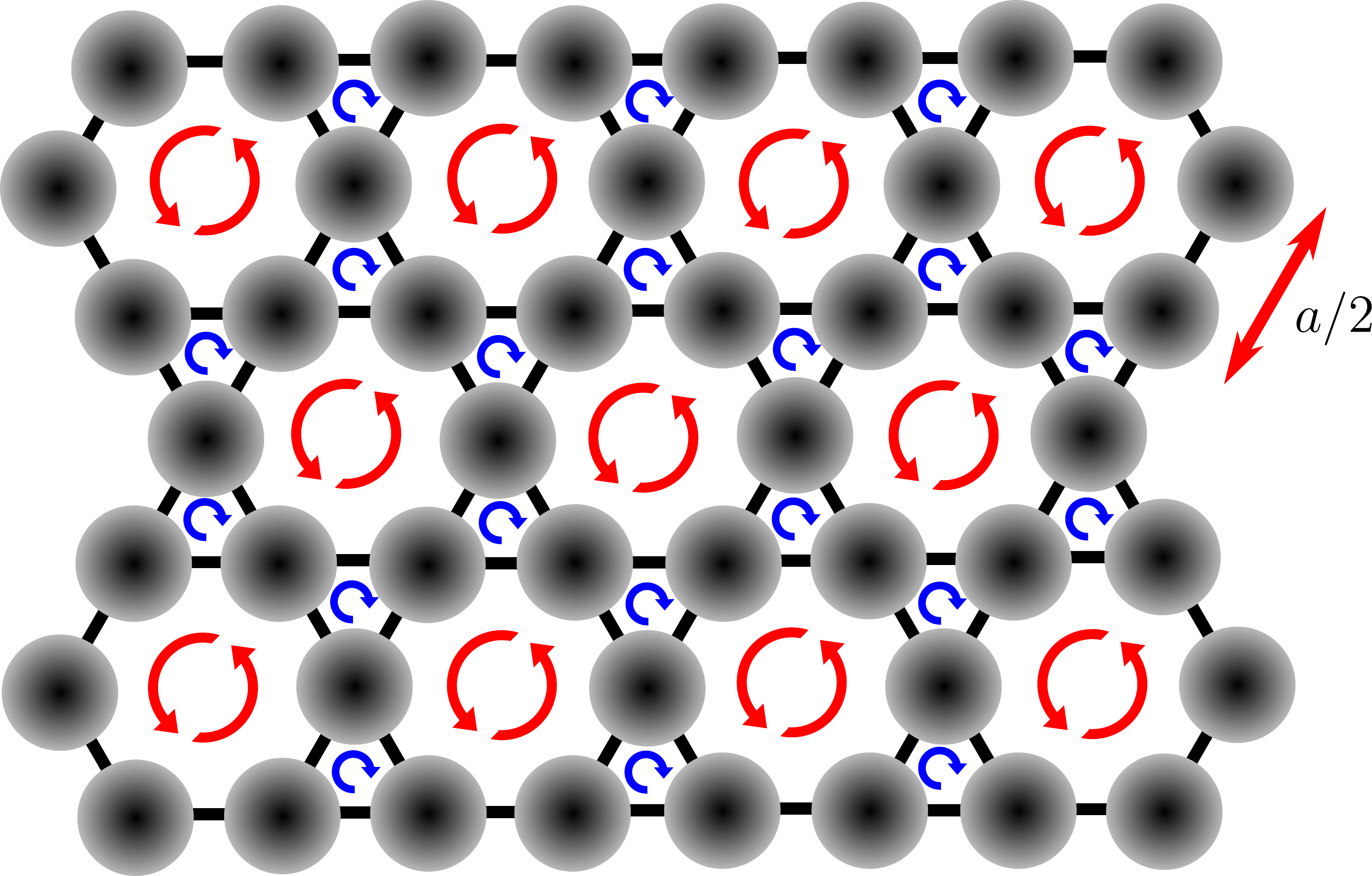}
	\caption{(Color online) A kagome lattice of Gaussian pump spots (gray) in the nonresonant excitation scheme. Tuning the width and distance between the spots we find that a stable vortex lattice forms spontaneously with periodic arrangement of charge $\pm2$ (red) and $\mp1$ (blue) vortices. Here we set $a = 16\mu$m and FWHM of the spots to $\sigma' = 0.31$.}
\label{fig0}
\end{figure}

{\bf Non-resonant Pumping.---} Several experimental~\cite{Tosi2012,Boulier2016} and theoretical studies~\cite{Keeling2008,Rodrigues2014} have considered the generation of vortex-antivortex lattices, their stability~\cite{Ma2017b} and the phase locking of vortices~\cite{Sigurdsson2014,Ma2017}. Here, we find that non-resonant excitation with a pumping field with intensity pattern corresponding to a kagome lattice spontaneously forms a lattice of vortices in phase (see Fig.~\ref{fig0} for a graphical representation of the excitation scheme).

In contrast to chiral pumping schemes~\cite{Dall2014}, our pumping field is non-chiral, such that the handedness of the vortices (that is, the sign of their winding number) is spontaneously chosen. All properties derived from the handedness of the vortices, such as an emergent non-trivial topology, are thus also spontaneously developed.

\begin{figure}[!t]
  \centering
	\includegraphics[width=\columnwidth]{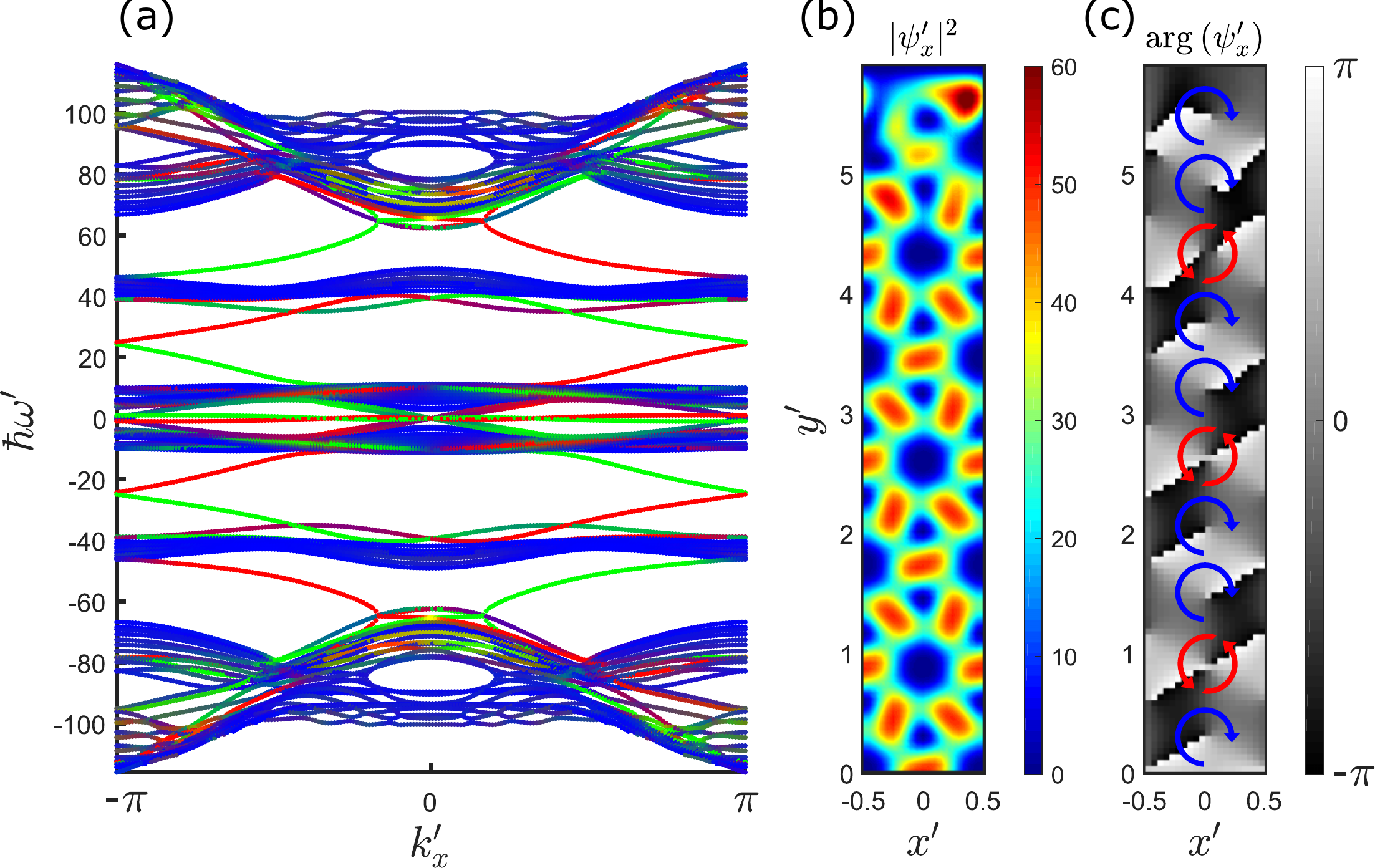}
	\caption{(Color online) (a) Bogoliubov spectrum of the nonresonantly driven polariton condensate in a strip geometry. The rotation of the vortex lattice in the condensate sets the topology of different bulk bands (blue) which are bridged by green (red) chiral edge states on the upper (lower) edge. (b) Density of the condensate vortex lattice in the upper half the of strip. (c) Phase of the condensate. Red double arrows indicate charge 2 vortices whereas blue single arrows charge -1 vortices. Parameters were set to: $a  =16\mu$m, $m=5\times10^{-5} m_0$, $\Gamma'_x=22$, $\Gamma_y' = 24$, $E_x=E_y=0$, $\alpha' |\psi_x'|^2 \approx 450$, $\Gamma_{NL}'  |\psi_x'|^2 \approx 50$, $g = 0.6$, and $P' \approx 88$ where $P'$ is the pump profile maximum. $E_p' \approx 83$ is set by the chemical potential of the $\psi_x$ condensate.}
\label{fig1}
\end{figure}

Given $\psi_x$, the calculation of the dispersion of the $y$-polarized field follows from Eq.~\ref{eq:Bogoliubov} and application of the Bloch theory. Considering a strip geometry (infinite in the $x$-direction, but with finite size in the $y$-direction), we show in Fig.~\ref{fig1}a the clear signature of topologically protected edges states in the $\psi_y$ component (green and red) separated by bulk bands. Fig.~\ref{fig1}b-c shows the steady state in the $\psi_x$ component. Setting $\Gamma_y > \Gamma_x$ and slowly ramping the pump intensity produces the $x$-polarized condensate.

The observed topological bandgap has a size on the order of $30\epsilon$. In real units, a typical polariton mass $m=5 \times10^{-5} m_0$, where $m_0$ is the mass of a free electron, and a typical lattice constant $a=16\mu$m sets an energy scale $\epsilon=3\mu$eV. The typical topological gap size is then on the order of $0.09$meV, which exceeds the decay rate and typical strengths of disorder in modern samples.

The presence of a hard-wall (Dirichlet) boundary at the edges of the strip in the $y'$-direction causes a static deformation of the $x-$component of the condensate at the edges but doesn't destabilize the lattice. In the current scheme where $\psi_y\sim0$ the vortex lattice is stable with zero net-interactions due to the cancellation between the $U_0$ and $U_1$ terms allowing us to freely play with $\alpha$ when resolving the Bogoliubov dispersion. In the case of an imbalance between $U_0$ and $U_1$, we have also found stable vortex-antivortex lattices, however there are limitations to their intensity as the vortices can become unstable with increasing pumping power~\cite{Sigurdsson2014}.

{\bf Resonant Pumping.---} Under resonant pumping,the phase of an incident optical field can be imprinted onto the polariton field, giving a more direct control. Considering an incident field $F(\mathbf{x})$ of the form of a kagome vortex-antivortex lattice, it is then straightforward to obtain a similar phenomenology to Fig.~\ref{fig1} (details are given in Sec.~\ref{supp1}).

\begin{figure}[t!]
\includegraphics[width=\columnwidth]{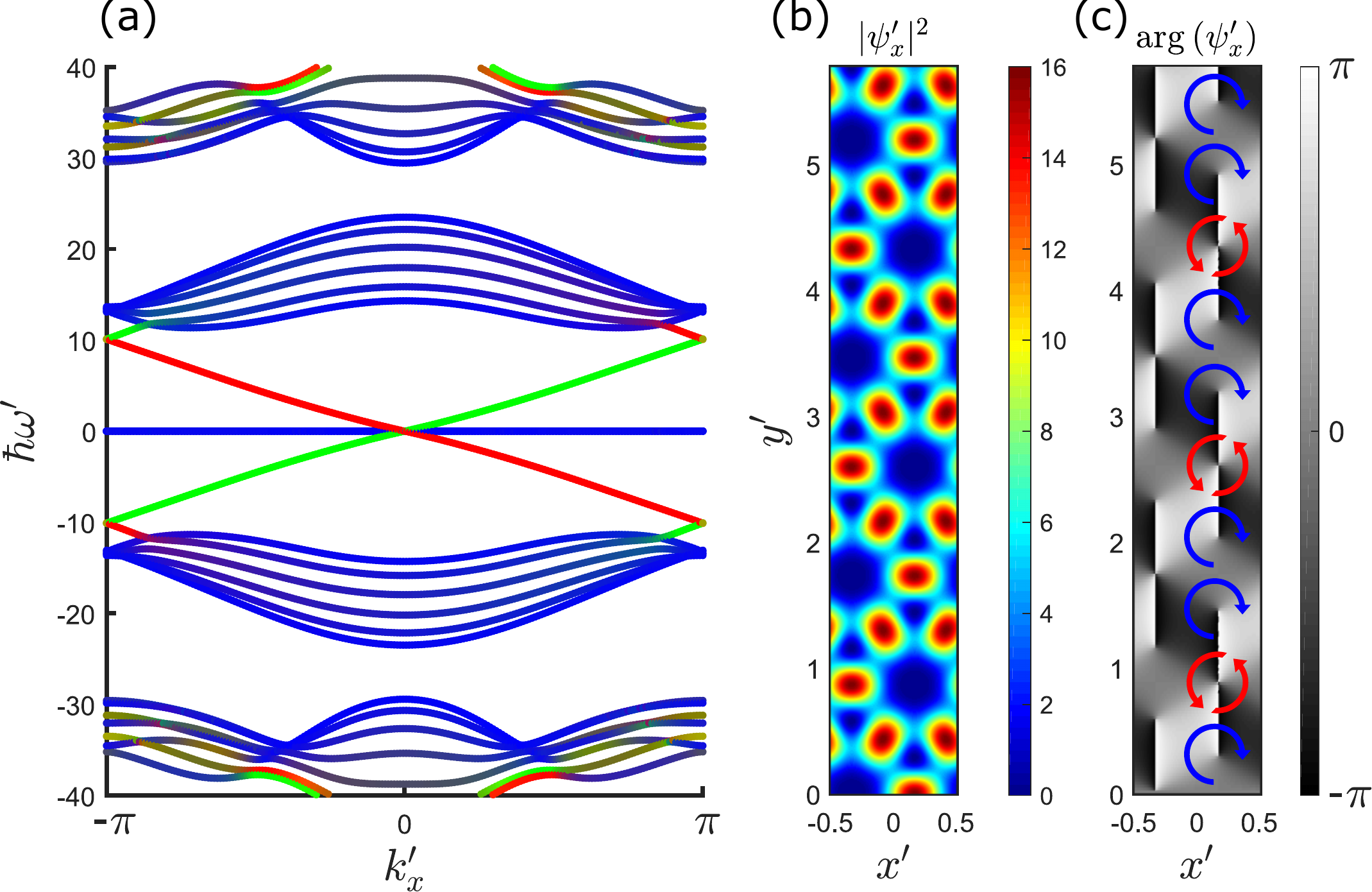}
\caption{(a) Optically-induced topological dispersion under resonant coherent pumping and kagome lattice potential $V(\mathbf{x})$. (b) Density of the condensate vortex lattice in the upper half the of strip. (c) Phase of the condensate. Red double arrows indicate charge 2 vortices whereas blue single arrows charge -1 vortices. Parameters: $\Gamma'=0.5$, $\alpha'|\psi'_x|^2=96$, $E_p'=0$, $V'=0$, $P(\mathbf{x})=0$, $\Gamma_\mathrm{NL}=0$.}
\label{fig:DispersionResonant}
\end{figure}

While topological behaviour is obtained in the case $V(\mathbf{x})=0$, it is still interesting to consider the case when there is an additional potential patterning of the microcavity. For simplicity, we assume that the potential is etched also in the form of a kagome lattice, $V'(\mathbf{x})=-\alpha'|\psi_x|^2$. We find that this allows the dispersion of the $y$-polarized component to not only be topological but also linear in $k$, as shown in Fig.~\ref{fig:DispersionResonant}.

It has been pointed out by several theoretical works in exciton-polariton systems that, due to the Landau criterion, a linear in $k$ dispersion corresponds to the phenomenon of superfluidity~\cite{Carusotto2004,Shelykh2006,Cancellieri2012,Carusotto2013}. The phenomenon was reported by several experimental works~\cite{Amo2009,Sanvitto2010,Lerario2017}, but never in the presence of a non-trivial topology. The traditional method of distinguishing superfluidity in polariton systems, which is based on observing a suppression of scattering with a defect, might not be considered sufficient in our system as the topological protection of the chiral edge state already prevents scattering with disorder. We expect that topological polariton superfluids should be characterized by their linear dispersion.

{\bf Conclusion.---} We have introduced schemes of optically-induced topological polaritons, characterized by the formation of chiral edge states, making use of the action of a condensate in one linear polarization on the dispersion of polaritons in the other. The scheme is compatible with un-etched planar microcavities and does not require any significant spin-orbit coupling (TE-TM splitting) in the sample or applied magnetic field.

Under non-resonant excitation, the phase-orientation of a vortex-antivortex (kagome) lattice is spontaneously chosen and responsible for the topological behaviour. Exciton-polaritons thus exhibit a unique feature among topological photonic systems, where an initially topologically trivial state undergoes a spontaneous topological phase transition with spontaneously chosen chirality. Under resonant excitation, the topological dispersion can be modified by the presence of an additional potential to obtain a dispersion linear in $k$, representative of a polariton superfluid.

{\bf Acknowledgements.---} H.S. acknowledges support by the Research Fund of the University of Iceland, The Icelandic Research Fund, Grant No. 163082-051, and the Icelandic Instruments Fund. G.L. acknowledges the EPSRC Programme on Hybrid Polaritonics for financial support. T.L. acknowledges support from the Ministry of Education (Singapore) grant 2015-T2-1-055.

\bibliography{bibliography}

\appendix
\begin{widetext}

\section{Topological dispersion in an unpatterned microcavity ($V=0$) under resonant pumping} \label{supp1}
We consider a patterned optical pump $F_x(\mathbf{x})$ taking the form of a kagome lattice, represented as a superposition of six plane waves:
\begin{equation}
F_x(\mathbf{x})=F_0\sum_{n=1}^6 e^{i(\mathbf{k}_n.\mathbf{x}+\phi_n)}
\end{equation}
where $F_0$ defines the amplitude; the wavevectors are $\mathbf{k}_{1,2}=k_0(\pm\sqrt{3}/2,1/2)$, $\mathbf{k}_2=k_0(\sqrt{3}/2,1/2)$, $\mathbf{k}_{3,4}=k_0(0,\pm1)$, and $\mathbf{k}_{5,6}=k_0(\pm\sqrt{3}/2,-1/2)$, where $k_0=4\pi/(\sqrt{3}a)$; and the phase factors are $\phi_{1,2,3,4}=0$ and $\phi_{5,6}=\pm2\pi/3$. Under resonant pumping, the terms $P(\mathbf{x})$ and $\Gamma_\mathrm{NL}$ are typically neglected in the modelling of exciton-polariton systems. Under such conditions, it is straightforward to show (for example by writing Eq.~3 in reciprocal space) that the polariton field $\psi_x(\mathbf{x})$ will adopt the intensity and phase structure of $F_x(\mathbf{x})$ in the stationary regime. The dispersion of the $y$-component is then obtained from application of the Bloch theory, giving the result shown in Fig.~\ref{fig:DispersionResonant2}. One sees a clear gap in the dispersion, where bulk states do not appear. The gap is topological, being bridged by a pair of chiral edge states that are localized on opposite edges of the strip.\\ \\
\indent This scheme may appear similar to that considered in Ref.~\cite{Bardyn2016}, which was based on a similar equation to Eq.~5, but the context and interpretation is very different. Here, by exciting a field with one linear polarization ($x$), we find that topological behaviour appears in the opposite linear polarization ($y$). It should be noted that the dispersion of $y$-polarized polaritons can be distinguished from the more highly populated $x$-polarization using polarization filtering. In addition, since the problem has been effectively divided into two parts, that is, solution of the $x$-polarized field from Eq.~3, and the dispersion of the $y$-polarized field from Eq.~5, it becomes easier to find parameters that give a topological bandgap. In particular the dispersion shown in Fig.~\ref{fig:DispersionResonant2} depends on the scaled parameter $\alpha'|\psi'_x|^2=160$ and shows a topological gap of typical size $\sim5\epsilon$.
\\
\begin{figure}[h!]
\includegraphics[width=0.7\columnwidth]{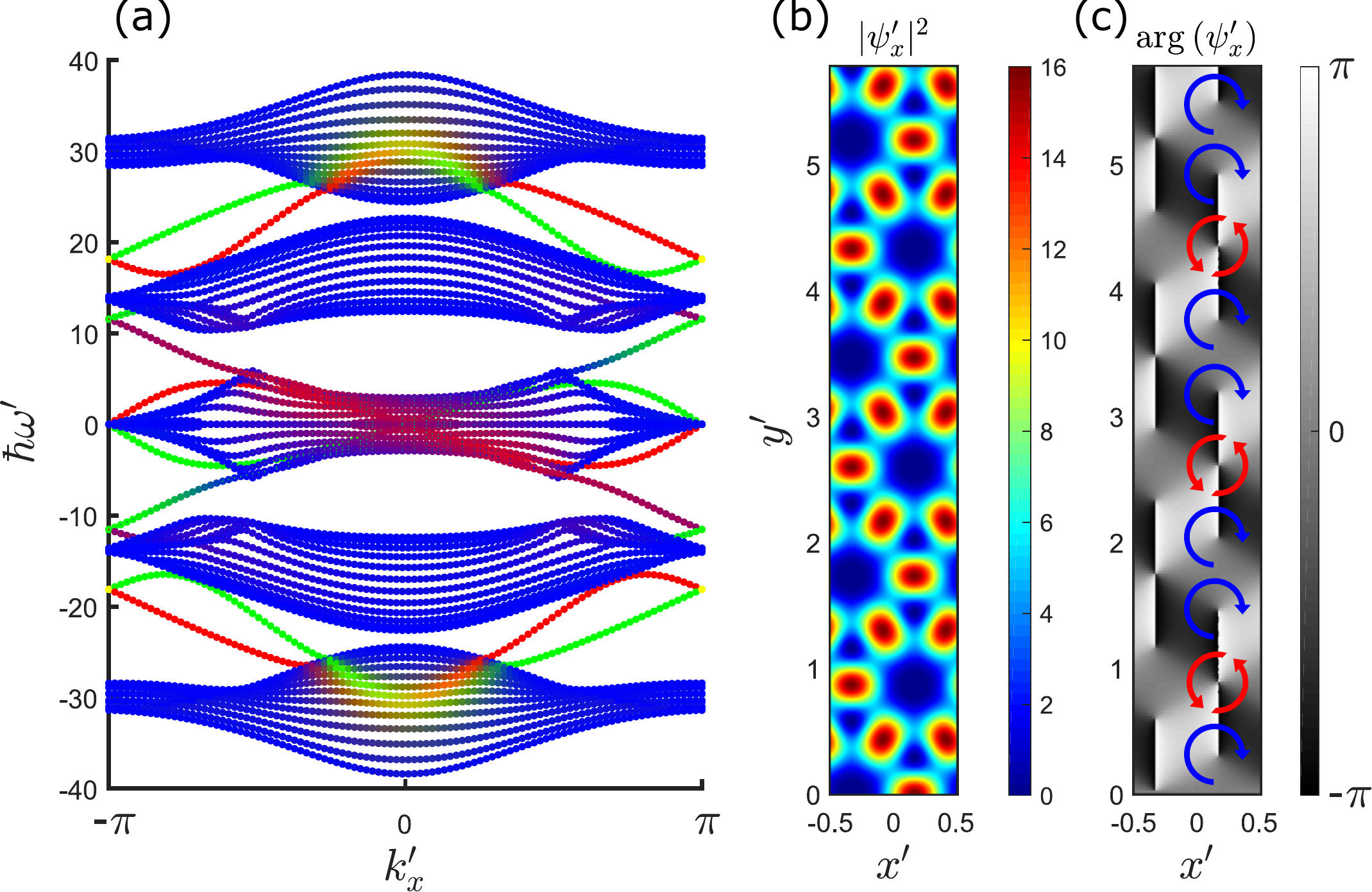}
\caption{(a) Optically-induced topological dispersion under resonant coherent pumping. (b) Density of the condensate vortex lattice in the upper half the of strip. (c) Phase of the condensate. Red double arrows indicate charge 2 vortices whereas blue single arrows charge -1 vortices. Parameters: $\Gamma'=0.5$, $\alpha'|\psi'_x|^2=160$, $E_p'=41$, $V'=0$, $P(\mathbf{x})=0$, $\Gamma_\mathrm{NL}=0$.}
\label{fig:DispersionResonant2}
\end{figure}

\end{widetext}

\end{document}